\documentclass{svproc}
\usepackage{url}
\usepackage{gensymb}
\usepackage{cite}
\usepackage{hyperref}
\usepackage{graphicx}
\usepackage{amsmath}
\usepackage{subcaption}
\usepackage{multicol}
\usepackage{float}
\usepackage{tikz}
\usepackage[affil-it]{authblk}

\usepackage{lineno}

\begin{document}
\mainmatter              
\title{Inclusive photon production at
	forward rapidities using PMD in p\textendash Pb collisions at $\sqrt{\it{s}_{\rm{NN}}} = 8.16$ TeV with ALICE}
\titlerunning{Photon multiplicity in p\textendash Pb collisions at $\sqrt{\it{s}_{\rm{NN}}} = 8.16$ TeV }  
%

\author{Mohammad Asif Bhat\\ \vspace{-0.3cm} $ ( $\rm{for the ALICE collaboration})}

\authorrunning{Mohammad Asif Bhat et al.} 
%
%

\institute{Bose Institute, EN 80, Sector V, Bidhan Nagar, Kolkata - 700091, West Bengal, India\\
\email{mohammad.asif.bhat@cern.ch}
\email{asifqadir@mail.jcbose.ac.in}
\email{bhatasif305@gmail.com}}

\maketitle              
\begin{abstract}
We report on the performance studies of two correction methods, namely the efficiency-purity method and the Bayesian unfolding method applied to the pseudorapidity distribution of photons. The pseudorapidity distribution  of inclusive photons at forward rapidities in the range \mbox{\small $(2.3<\eta<3.9)$} in p\textendash Pb collisions at $\sqrt{\it{s}_{\rm{NN}}} = 8.16$ TeV is obtained with the HIJING Monte Carlo event generator.
The simulated data samples were obtained from the Photon Multiplicity Detector (PMD) in ALICE.
\end{abstract}
\section{Introduction}
Relativistic heavy-ion collisions are believed to produce a Quark Gluon Plasma (QGP)~\cite{Aamodt:2008zz} which hadronizes dominantly into pions, of which approximately one third are neutral pions that decay into photons. Therefore, the measurement of photon multiplicity can provide important information about bulk physics~\cite{Adam:2015gka} from initial scatterings to the final state QGP effects. The measurement of photon multiplicity in p\textendash Pb collisions~\cite{Adare:2013lkk} is very important as it is an intermediate step going from hadronic to heavy-ion collisions.\par

In this analysis we generated the pseudorapidity distribution of inclusive photons at forward rapidities ($2.3 < \eta < 3.9$) in p\textendash Pb collisions at $\sqrt{\it{s}_{\rm{NN}}} = 8.16$ TeV using the HIJING~\cite{Wang:1991hta} Monte Carlo event generator. The photons are detected by the Photon Multiplicity Detector (PMD)~\cite{Dellacasa:1999ar,Adams:2005aa}. The raw pseudorapidity distribution of inclusive photons was corrected by following two methods: (i) Efficiency-purity method (ii) Bayesian unfolding method.
 We used the minimium bias events, which produce at least one hit in V0A and V0C trigger detectors.
The events with the collision vertex coordinate along the beam axis {\it{z}} lying between -10 to +10 cm from the nominal interaction point are taken for physics analysis.


%
\section{Pseudorapidity distribution of photons}
\begin{figure}[htbp!]\vspace{-0.6cm}
	\begin{center}
		\begin{minipage}{0.49\textwidth}
			\centering
			\includegraphics[width=1\linewidth]{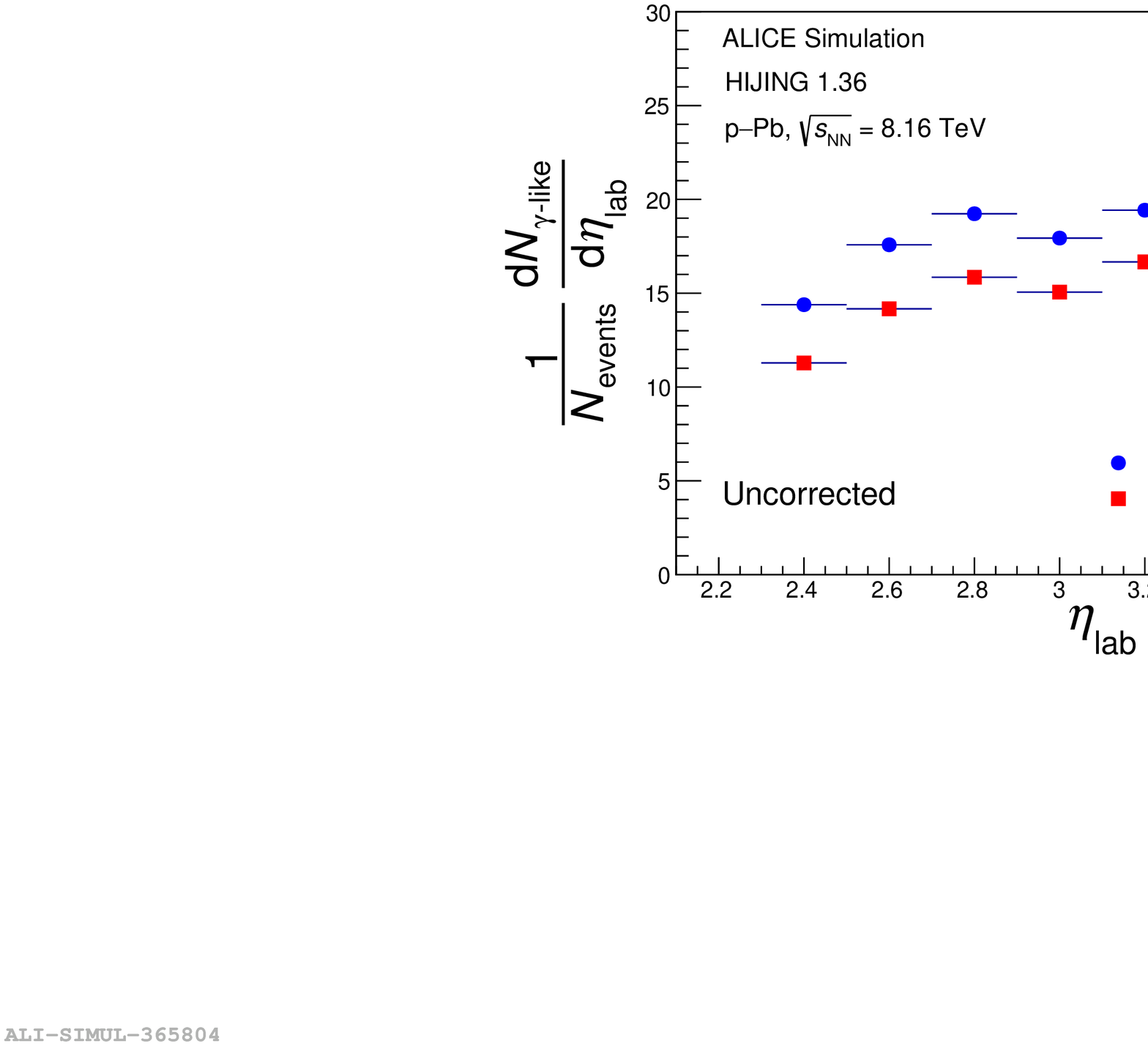}
		\end{minipage}
		\begin{minipage}{0.49\textwidth}
			\centering
			\includegraphics[width=1\linewidth]{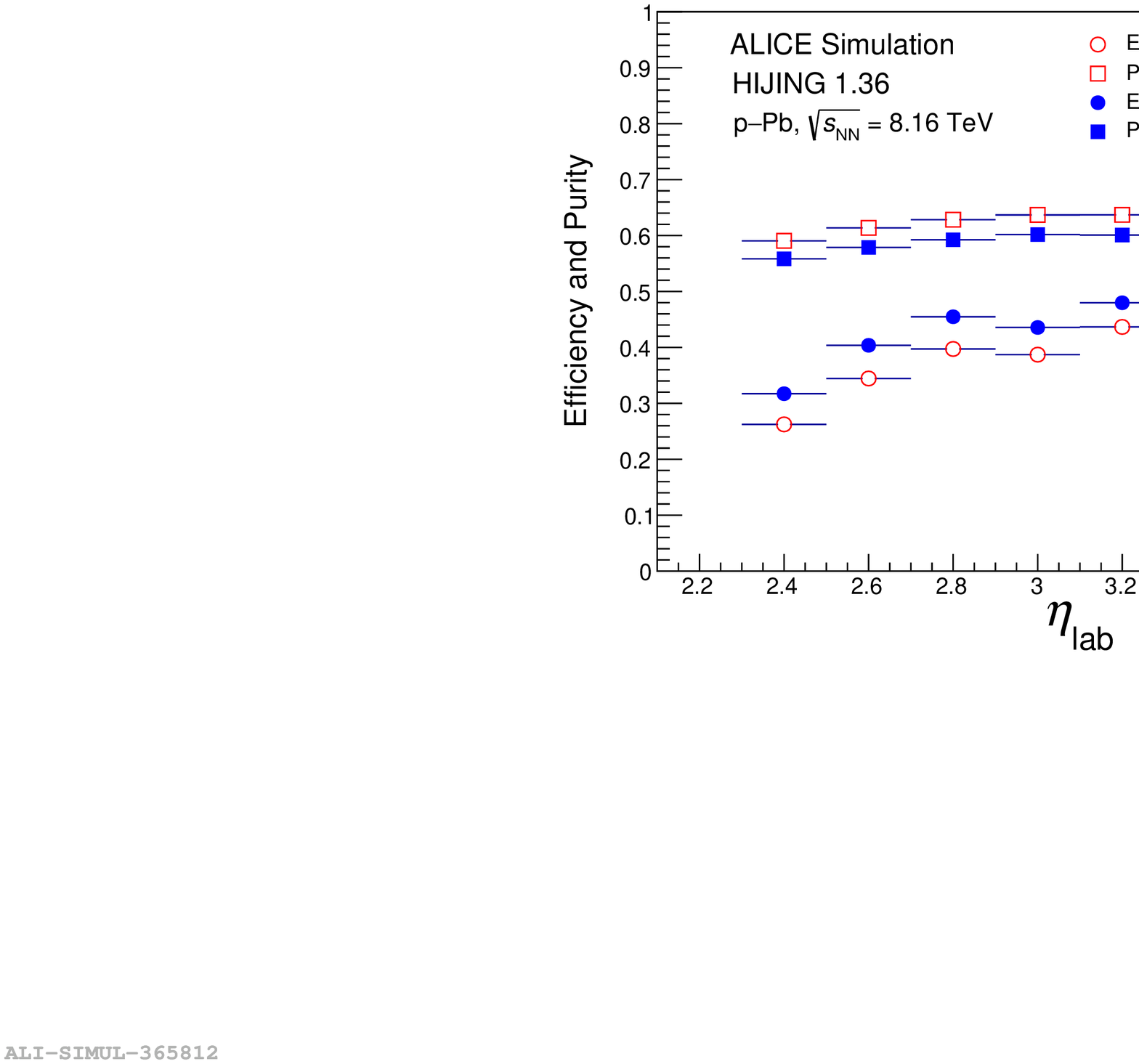}
		\end{minipage}
		\caption{\small Left: Uncorrected pseudorapidity distribution of photons in p\textendash Pb collisions at $\sqrt{\it{s}_{\rm{NN}}} = 8.16$ TeV using MC event generator HIJING. Right: Efficiency and purity values as a function of pseudorapidity.}
		\label{1} 
	\end{center}
\end{figure}

Photons are identified using two different conditions imposed on all detected clusters on the PMD. These are \small{ADC $>$ 6MPV,\mbox{$N_{cell}>2$}} and \small{ADC $>$ 9MPV,\mbox{$N_{cell}>2$}},
where ADC is the sum of the ADC channel numbers of the cells within the cluster,  MPV = 72 ADC channels is the most probable value of the pion ADC distribution and $N_{cell}$ is the average number of cells affected by 3 GeV pion from the test beam result~\cite{ALICE:2014rma}. The pseudorapidity distribution of photons is shown on the left side of Fig~\ref{1} . This distribution is affected by 
 the detector finite resolution, limited detector efficiency and acceptance and secondary contamination such as hadron clusters. Therefore correction is needed to be applied to get the true distribution. We used two methods for correction as detailed in the following section.
\vspace{-0.8cm}
\section{Correction methods}
\subsection{ Efficiency-purity method}
Efficiency ($\epsilon_{\gamma}$) is defined as the ratio of the number of detected photons to the number of incident photons within the same acceptance coverage
\begin{equation}
\epsilon_{\gamma} \;\; = \;\;  \frac{N_{\gamma-detected}}{N_{\gamma-incident}}\;\;\;,
\end{equation}
where $\gamma-detected$  are those $\gamma-like$ clusters whose cluster identity matches with the incident photon identity, $\gamma-incident$ are the incident photons produced from HIJING. \\
Purity $(\it{p})$ is defined as the ratio of the number of detected photons to the number of gamma-like clusters within the same acceptance coverage
\begin{equation}
p   \;\;= \;\;  \frac{N_{\gamma-detected}}{N_{\gamma-like}}\;\;\;,
\end{equation}
where $\gamma-like$ or $\gamma-measured$ are those clusters that satisfy the photon-hadron discrimination conditions.\\
The right panel of Fig~\ref{1} \space shows the efficiency and purity values as a function of pseudorapidity
for two different photon-hadron discrimination conditions.\\
To obtain the corrected photon distribution we used the following correction factor: 
\begin{equation}
N_{\gamma-true}= \frac{p}{\epsilon_{\gamma}}\times N_{\gamma-measured}
\end{equation}

\subsection{Bayesian unfolding method}
Unfolding is a technique to get the true distribution from the measured distribution. 
Bayesian unfolding \cite{DAgostini:1994fjx} is based on the Bayes's theorem
\begin{equation}
P(A|B) = \frac{P(B|A)\times P(A)}{P(B)}\;\;\;\;.
\end{equation}
In our notation
\begin{equation}
\bar{A_{ij}}= \frac{A_{ji}f_{i}}{\varSigma_{k} A_{ik}f_{k}}
\end{equation}

\begin{equation}	
f_{i} = \varSigma_{j} \bar{A_{ij}} g_{j} 
\end{equation}\\
where $f_{i}$ is the unfolded distribution, $\bar{A_{ij}}$ is the inverse of the response matrix $A_{ij}$ and $g_{j}$ is the measured distribution.
The fluctuation in the error is controlled by the regularization parameters such as the number of iterations and smoothing. 
The unfolding was carried out in each pseudorapidity bin and for each pseudorapidity bin we have obtained the response matrix, measured ($\gamma-like$) distribution and the unfolded distribution for different set of regularization parameters. The sets of parameters which gives the minimium possible fluctuations in the error were selected. The average value of the unfolded distribution divided by the bin width was used to get the pseudorapidity distribution of photons bin by bin.
\begin{figure}[htbp!]\vspace{-0.6cm}
	\begin{center}
		\begin{minipage}{0.49\textwidth}
			\centering
			\includegraphics[width=1\linewidth]{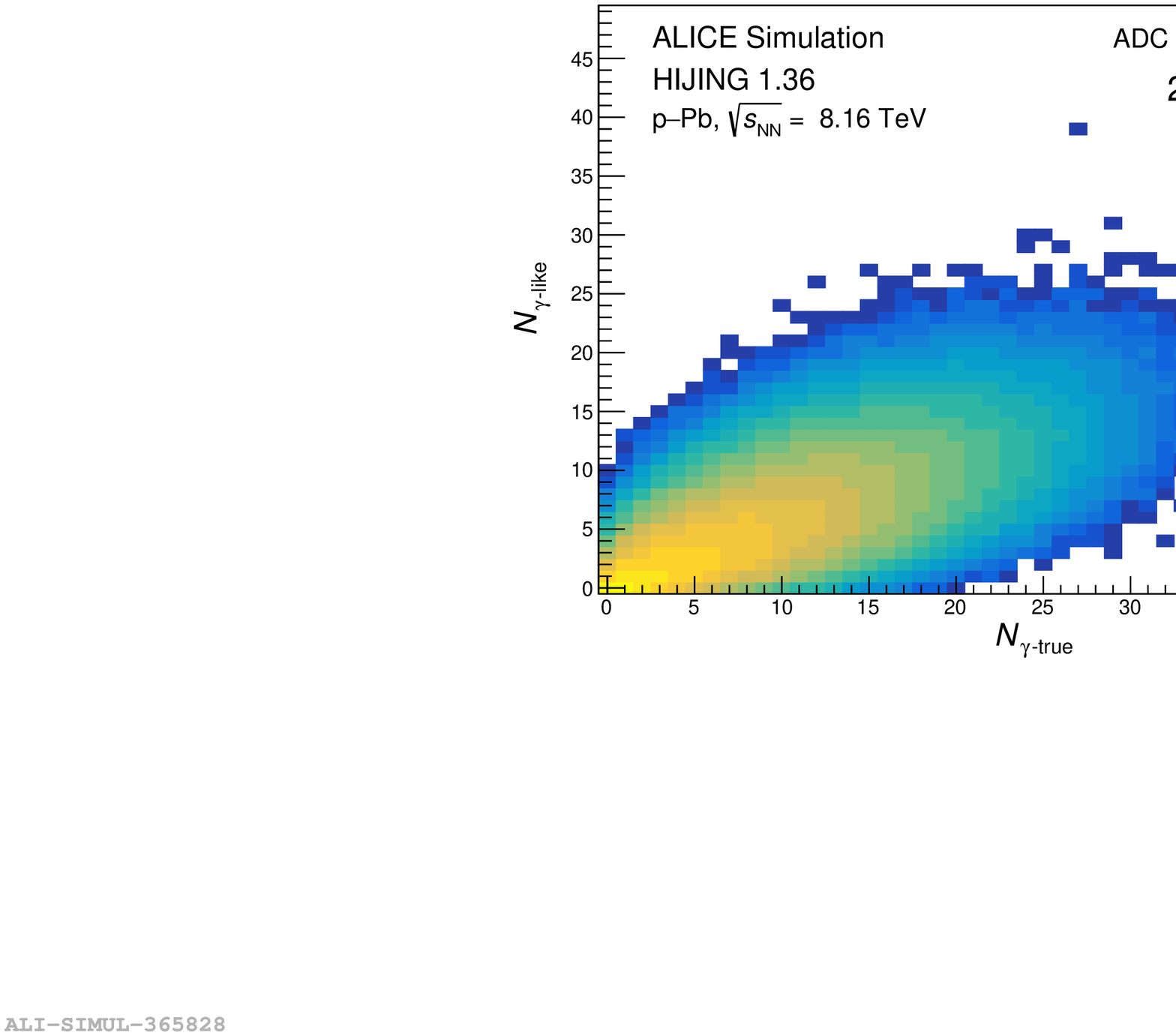}
		\end{minipage}
		\begin{minipage}{0.49\textwidth}
			\centering
			\includegraphics[width=1\linewidth]{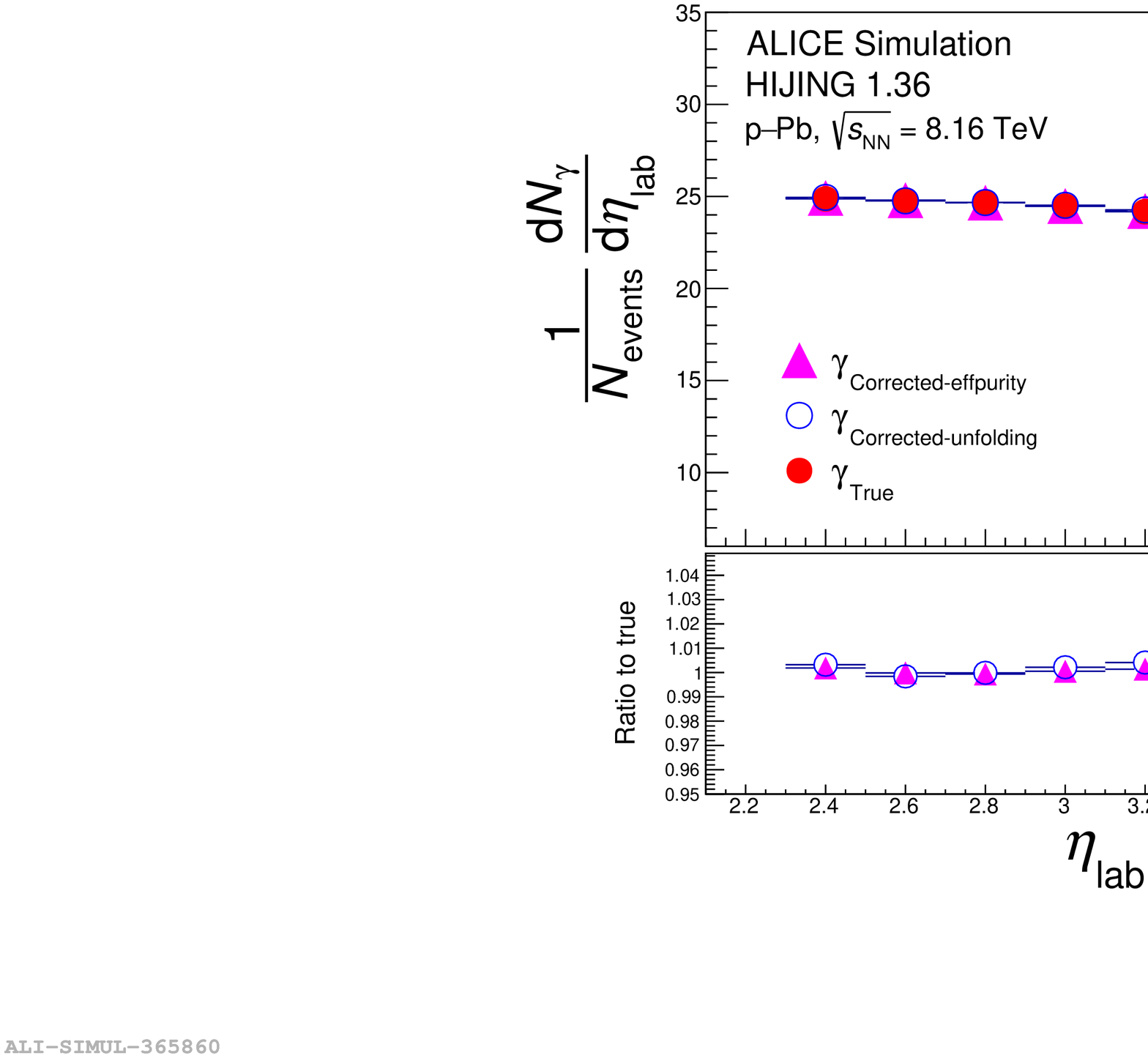}
		\end{minipage}
		\caption{\small Left: Response matrix for interval \mbox{$(2.3 < \eta < 2.5)$}.  Right: Efficiency-purity corrected and Bayesian-unfolded pseudorapidity distribution of photons compared with the pseudorapidity distribution of true incident photons from HIJING.}
		\label{2} 
	\end{center}
\end{figure}
\section{Results}
The pseudorapidity distribution of photons obtained after applying the two correction methods is
compared with the pseudorapidity distribution of true incident photons from HIJING as shown on right side of Fig~\ref{2} . The ratios of the true pseudorapidity  distribution with the one obtained after applying each of the two correction methods are consistent with unity with an error of 0.5\%. This shows that while analyzing the data any one of the two correction methods can be applied and the difference arising from using the other can be included in the systematic uncertainty.
\section{Summary}
We reported on the corrected pseudorapidity distribution of photons obtained by two correction methods: the efficiency-purity method and the Bayesian unfolding method. The corrected pseudorapidity distributions of photons obtained by these two correction methods are in good agreement with the pseudorapidity distribution of true incident photons from HIJING.
\bibliography{reference}

\end{document}